\documentclass[10pt,english,floatfix,nofootinbib,superscriptaddress,aps,prd,preprint, twocolumn]{revtex4}
\usepackage[utf8]{inputenc}
\usepackage{float}
\usepackage{array}
\usepackage{lipsum}
\usepackage{bbm}
\usepackage{dsfont}
\usepackage{graphicx}
\usepackage{amsmath}
\usepackage{tikz}
\usepackage{multirow}
\usepackage{braket}
\usepackage{bbm}
\usetikzlibrary{quotes,angles}
\usetikzlibrary{arrows} 
\usetikzlibrary{decorations.markings}
\usepackage{graphicx}
\usepackage[english]{babel}
\usepackage{color}
\usepackage{subfig}
\usepackage{caption}
\usepackage{tensor}
\usepackage{esint}
\usepackage[dvips]{epsfig}
\usepackage[dvips]{graphicx}
\usepackage{float}
\usepackage{units}
\usepackage{textcomp}
\usepackage{mathrsfs}
\usepackage{amsmath}
\usepackage[makeroom]{cancel}
\usepackage{amssymb}
\usepackage{amsbsy}
\usepackage{amsfonts}
\usepackage{amssymb,mathrsfs,xcolor}
\usepackage{esint}
\usepackage{array}
\usepackage{graphicx}
\usepackage{slashed}

\newcommand{\ie}{\begin{equation}}
\newcommand{\fe}{\end{equation}}
\newcommand{\se}{\begin{eqnarray}}
\newcommand{\ff}{\end{eqnarray}}

\usepackage{hyperref}
\hypersetup{colorlinks,breaklinks,
			citecolor=[rgb]{0,0.0,1.0},
            urlcolor=[rgb]{0.0,0.0,1.0},
            linkcolor=[rgb]{0,0.5,0.9}}

\begin{document}

\title{Effects of bumblebee gravity on neutrino motion}

\author{Yuxuan Shi}
\email{shiyx2280771974@gmail.com}
\affiliation{Department of Physics, East China University of Science and Technology, Shanghai 200237, China}

\author{A. A. Ara\'{u}jo Filho}
\email{dilto@fisica.ufc.br}
\affiliation{Departamento de Física, Universidade Federal da Paraíba, Caixa Postal 5008, 58051--970, João Pessoa, Paraíba,  Brazil.}
\affiliation{Departamento de Física, Universidade Federal de Campina Grande Caixa Postal 10071, 58429-900 Campina Grande, Paraíba, Brazil.}
\affiliation{Center for Theoretical Physics, Khazar University, 41 Mehseti Street, Baku, AZ-1096, Azerbaijan.}


\date{\today}

\begin{abstract}

This study explores how the spontaneous violation of Lorentz symmetry—modeled through a black hole solution in the context of bumblebee gravity—affects the propagation and dynamics of neutrinos. The investigation centers on three distinct aspects: the rate of energy deposition due to neutrino--antineutrino pair annihilation, modifications to the neutrino oscillation phase driven by the underlying spacetime structure, and the influence of gravitational lensing on flavor conversion probabilities. To support the theoretical considerations, numerical simulations are conducted for oscillation probabilities in a two--flavor framework, taking into account both normal and inverted mass orderings. For comparison, the outcomes are juxtaposed with those obtained in a different Lorentz--violating background, namely, a black hole solution within Kalb--Ramond gravity.

\end{abstract}

\maketitle


\section{Introduction}

Lorentz symmetry—the invariance of physical laws under transformations between inertial reference frames—has been extensively supported by experimental evidence. Nonetheless, several theoretical models suggest that this symmetry might not hold in extreme energy scales or in regions with intense gravitational fields. Numerous investigations have examined the implications of such symmetry violations within diverse theoretical settings \cite{7,4,heidari2023gravitational,3,1,araujo2023thermodynamics,8,2,6,ghosh2023does,5}. These violations are generally classified into two types: spontaneous and explicit \cite{bluhm2006overview}. In the explicit case, Lorentz--breaking contributions are inserted directly into the field equations, which can generate measurable directional dependencies. Conversely, spontaneous Lorentz violation arises when the field equations themselves remain Lorentz invariant, but the vacuum solution breaks the symmetry, allowing the violation to manifest through the properties of the ground state \cite{bluhm2008spontaneous,colladay1998lorentz,kostelecky2004gravity}.

Growing interest has been directed toward modified gravity theories that feature spontaneous Lorentz symmetry violation, especially within the framework of the Standard Model Extension \cite{13,heidari2024scattering,AraujoFilho:2024ykw,KhodadiPoDU2023,11,liu2024shadow,filho2023vacuum,9,10,12}. A prominent example of such models is the bumblebee theory, where a dynamical vector field develops a nonzero vacuum expectation value, establishing preferred directions in spacetime and thereby inducing a local breakdown of Lorentz invariance. This violation does not arise from explicit terms in the fundamental equations but instead emerges through the properties of the vacuum itself \cite{bluhm2006overview}. Moreover, Bumblebee models have been applied to a wide range of gravitational scenarios, focusing particularly on their impact on the thermodynamic aspects \cite{paperrainbow,anacleto2018lorentz,reis2021thermal,aa2021lorentz,araujo2021higher,araujo2022does,araujo2021thermodynamic,araujo2022thermal}.

The formulation of static and spherically symmetric black hole solutions within bumblebee gravity was first introduced in Ref.~\cite{14}, prompting a broad spectrum of research into the physical consequences of such backgrounds. Schwarzschild--like metrics modified by spontaneous Lorentz violation have subsequently been applied to diverse scenarios, including the impact on Hawking radiation emission processes \cite{kanzi2019gup}, the gravitational lensing deviations induced by the altered geometry \cite{15} and the study of matter infall and accretion dynamics near compact objects \cite{17,18}. Investigations have also addressed how these deformations affect the quasinormal mode structure of perturbations \cite{19,Liu:2022dcn}. More recently, the role of such symmetry--breaking backgrounds in quantum processes—especially in particle production near black hole horizons—has been analyzed for vector \cite{araujo2025does} and tensor field \cite{AraujoFilho:2024ctw} scenarios.

The search for generalized black hole solutions beyond the standard (A)dS--Schwarzschild spacetime has led to a range of theoretical extensions, including scenarios in which the conventional vacuum assumptions are relaxed, resulting in modified geometric and dynamical properties of spacetime \cite{20}. A notable refinement of this idea arises in the context of bumblebee gravity, where a vector field spontaneously acquires a non--zero expectation value in the temporal direction. This mechanism triggers the breaking of local Lorentz invariance and yields new possibilities beyond general relativity \cite{24,21,23,lambiase2023probing,22,ovgun2019exact,Magalhaes:2025nql,marques2023braneworlds,engels2016bumblebee,kanzi2021greybody,mangut2023probing,mangut2025lorentz,Magalhaes:2025lti}.

Although quark mixing and meson oscillations are also well established, neutrinos are unique because their tiny masses and flavor–mass misalignment allow coherent oscillations to persist over macroscopic and even astrophysical distances \cite{Pontecorvo1,Pontecorvo2,maki1962remarks}. Owing to this property, they have become especially sensitive probes of new physics and gravitational effects \cite{neu42,neu43,neu44}. One of their most notable features is that the states responsible for weak interactions—referred to as flavor states—do not coincide with the states that possess definite mass. This misalignment causes neutrinos to evolve as a mixture of mass eigenstates, leading to an oscillatory pattern in their flavor composition as they propagate through space \cite{neu40,neu39,neu41}. The resulting flavor change, known as neutrino oscillation, arises from quantum interference between these mass components.

In a flat spacetime framework, neutrino oscillations are determined by the differences between the squares of the neutrino mass eigenvalues, rather than the individual mass magnitudes. These mass--squared differences are defined as
$$
\Delta m^2_{ij} = m_i^2 - m_j^2,
$$
with commonly analyzed quantities being $|\Delta m^2_{21}|$, $|\Delta m^2_{31}|$, and $|\Delta m^2_{23}|$. Since the transition probabilities depend exclusively on these differences, oscillation measurements are inherently insensitive to the absolute scale of neutrino masses, providing information only about their relative gaps \cite{neu45}.

Gravitational fields can markedly influence neutrino oscillation dynamics, introducing effects absent in flat spacetime scenarios. While in flat geometry the flavor transitions depend solely on differences between mass--squared values, curved spacetime geometries modify the oscillation phase in a way that can reveal sensitivity to the absolute mass scale of neutrinos \cite{Chakrabarty:2023kld,Alloqulov:2024sns,Shi:2024flw,AraujoFilho:2025rzh,Shi:2025rfq}. These curvature--induced corrections alter the conventional phase relations and become especially significant for highly energetic neutrinos emitted from astrophysical or cosmological sources \cite{Shi:2023kid}. As neutrinos propagate through regions with nontrivial geometry, the accumulated phase shift encodes information about both their internal parameters and the structure of the spacetime they traverse. Observational discrepancies between expected flavor distributions and those predicted by standard oscillation models can thus indicate gravitational influences. In this manner, this approach provides a pathway to explore the mass spectrum of neutrinos while simultaneously probing the nature of the gravitational environments encountered during their journey \cite{neu53,neu46,neu49,neu48,neu47,neu50,neu51,Shi:2023hbw,neu52}.

Neutrino flavor dynamics are strongly shaped by the geometry of the spacetime they traverse. When described in geometric terms, the phase acquired along their trajectories reflects the influence of the gravitational background \cite{neu54}, as one should naturally expect. In highly curved environments—such as near compact massive bodies—gravitational lensing can distort neutrino paths, causing them to converge or intersect. This bending alters the interference between different mass eigenstates, leading to modifications in the oscillation behavior and affecting the resulting flavor transition probabilities \cite{Shi:2025rfq,neu53,Shi:2024flw}.

Recent investigations have increasingly focused on how spacetime curvature—particularly due to gravitational lensing—can alter the quantum coherence necessary for neutrino flavor transitions \cite{neu57,neu58,neu56}. In scenarios where the gravitational source exhibits rotation, the complexity of the oscillation process increases substantially. Rotational effects, as shown by Swami, introduce corrections to the phase evolution of neutrinos that depend on the angular momentum of the central mass. These corrections can lead to either amplification or suppression of oscillation probabilities, contingent on the nature of the spacetime through which the neutrinos propagate. Such rotational influences are especially relevant in gravitational fields associated with stars of solar mass scale \cite{neu59}.

Neutrino flavor transitions have also been explored within geometries that deviate from perfect spherical symmetry, particularly in axially deformed spacetimes. In these cases, a deformation factor denoted by $\gamma$ modifies the structure of static, asymptotically flat metrics. This geometric alteration affects the phase evolution of neutrinos as they propagate, potentially introducing a dependence on the absolute neutrino masses. Such mass sensitivity does not arise in conventional flat spacetime treatments, highlighting the impact of background geometry on oscillation dynamics \cite{neu60}.

This work examines the impact of Lorentz symmetry breaking—arising from a black hole configuration in bumblebee gravity—on neutrino behavior. Rather than focusing solely on propagation, the analysis addresses three interconnected phenomena: energy transfer resulting from neutrino–antineutrino annihilation, phase distortions in neutrino oscillations due to spacetime curvature, and alterations in flavor transition rates caused by lensing effects. The study employs a two--flavor oscillation model to perform numerical evaluations under both normal and inverted mass orderings. Additionally, the findings are contrasted with those derived from an alternative symmetry--violating framework based on the Kalb--Ramond black hole geometry, allowing a broader assessment of how different Lorentz--violating backgrounds affect neutrino dynamics.

\section{The bumblebee solution in the metric formalism}

Our analysis focuses on the bumblebee black hole solution formulated within the metric formalism \cite{14}
\begin{align}
\label{mmainin}
\mathrm{d}s^{2} = & -\left(1-\dfrac{2M}{r}\right)\mathrm{d}t^{2} + (1+\ell)\left(1-\dfrac{2M}{r}\right)^{-1}\mathrm{d}r^{2}\\
& + r^{2}\mathrm{d}\theta^{2} + r^{2} \sin^{2}\mathrm{d}\varphi^{2}.
\end{align}

Recent developments have explored the bumblebee black hole solution in the metric formalism as a platform for addressing quantum and semiclassical effects, such as the degradation of entanglement near horizons \cite{Liu:2024wpa} and mechanisms of particle creation \cite{araujo2025does}. The model has also been adapted to cosmological settings, including connections with anisotropic Kasner--type geometries \cite{Neves:2022qyb}, and has been applied to describe compact stellar configurations \cite{Neves:2024ggn}. Modifications in gravitational wave polarization patterns were investigated within this framework as well \cite{Liang:2022hxd}, while the quasinormal mode spectrum has been extensively analyzed in perturbative regimes \cite{Oliveira:2021abg}.

From a geometric perspective, extensions involving topological defects \cite{Gullu:2020qzu} and non--commutative particle dynamics \cite{KumarJha:2020ivj} have broadened its applicability. Variants of the solution including a cosmological constant \cite{20}, as well as those mimicking Kerr--like rotation \cite{Ding:2019mal,Liu:2019mls}, have been constructed and studied in detail. Investigations of gravitational lensing via both the Gauss--Bonnet approach \cite{Ovgun:2018ran} and geodesic-based methods \cite{Li:2020wvn}, along with circular orbit analyses, have enriched the phenomenological profile of the solution.

The thermodynamic behavior of bumblebee black holes, particularly under corrections from the generalized uncertainty principle \cite{Gomes:2018oyd,Ovgun:2019ygw,Kanzi:2019gtu}, has been a major topic of study. Additional research has focused on Ricci dark energy scenarios \cite{Jesus:2019nwi} and matter accretion processes \cite{Yang:2018zef}, reinforcing the model’s relevance in both fundamental and observational contexts.

The next section will be investigating the influence of the Bumblebee black hole background, as formulated in Ref.~\cite{14}, on multiple facets of neutrino behavior. The analysis opens with the evaluation of the energy release associated with neutrino--antineutrino pair annihilation near the black hole. Attention is then directed toward how the underlying spacetime structure alters the oscillation phase and reshapes the flavor transition probabilities. Building on these elements, the study proceeds to assess the role of gravitational deflection on neutrino propagation. Finally, numerical simulations are provided to illustrate the impact of Lorentz symmetry breaking on neutrino dynamics.


\section{Annihilation--induced energy deposition rate.}

This segment focuses on the process of energy deposition occurring in a modified gravitational background characterized by the Lorentz--violating coefficient $\ell$, as introduced in Eq.~(\ref{mmainin}). Within this approach, the primary mechanism for energy transfer stems from the annihilation of neutrinos and antineutrinos. The rate at which energy is deposited—quantified per unit time and volume—is given by the expression \cite{Salmonson:1999es}:
\ie
\dfrac{\mathrm{d}\mathrm{E}(r)}{\mathrm{d}t\mathrm{d}V}=2 \, \mathrm{K} \,\mathrm{G}_{f}^{2}\, \mathfrak{f}(r)\iint
\Tilde{n}(\varepsilon_{\nu})\Tilde{n}(\varepsilon_{\overline{\nu}})
(\varepsilon_{\nu} + \varepsilon_{\overline{\nu}})
\varepsilon_{\nu}^{3}\varepsilon_{\overline{\nu}}^{3}
\mathrm{d}\varepsilon_{\nu}\mathrm{d} \varepsilon_{\overline{\nu}}
\fe
with
\ie
\mathrm{K} = \dfrac{1}{6\pi}(1\pm4\sin^{2}\theta_{W}+8\sin^{4} \theta_{W}).
\fe

By adopting the standard value $\sin^{2}\theta_{W} = 0.23$ for the Weinberg angle, it becomes possible to derive explicit formulas for the energy deposition rates corresponding to different neutrino--antineutrino annihilation channels. These relations, as obtained in \cite{Salmonson:1999es}, encapsulate how the weak interaction coupling and the flavor composition of the annihilating particles influence the overall rate of energy transfer
\ie
\mathrm{K}(\nu_{\mu},\overline{\nu}_{\mu}) = \mathrm{K}(\nu_{\tau},\overline{\nu}_{\tau})
=\dfrac{1}{6\pi}\left(1-4\sin^{2}\theta_{W} + 8\sin^{4}\theta_{W}\right),
\fe
and
\ie
\mathrm{K}(\nu_{e},\overline{\nu}_{e})
=\dfrac{1}{6\pi}\left(1+4\sin^{2}\theta_{W} + 8\sin^{4}\theta_{W}\right).
\fe

The evaluation of energy release from each neutrino--antineutrino channel relies on the adopted value of the Weinberg angle, with $\sin^2\theta_W = 0.23$, which determines the coupling constants involved in weak interactions. The interaction strength is governed by the Fermi constant, set as $\mathrm{G}_f = 5.29 \times 10^{-44} , \text{cm}^2 , \text{MeV}^{-2}$. These inputs allow for the computation of distinct energy deposition rates associated with each flavor combination. After performing the angular integration, the resulting expression for the deposition contribution is obtained in the form presented in \cite{Salmonson:1999es}
\begin{align}
\mathfrak{f}(r)&=\iint\left(1-\bm{\Tilde{\Omega}_{\nu}}\cdot\bm{\Tilde{\Omega}_{\overline{\nu}}}\right)^{2}
\mathrm{d}\Tilde{\Omega}_{\nu}\mathrm{d}\Tilde{\Omega}_{\overline{\nu}}\notag\\
&=\dfrac{2\pi^{2}}{3}(1 - x)^{4}\left(x^{2} + 4x + 5\right)
\end{align}
with
\ie
x = \sin\theta_{r}.
\fe

At a given radial distance $r$, the angle $\theta_r$ quantifies how much a particle’s trajectory deviates from the tangential direction associated with a circular orbit at that radius. The directions of motion for neutrinos and antineutrinos are represented by the unit vectors $\Tilde{\Omega}_{\nu}$ and $\Tilde{\Omega}_{\overline{\nu}}$, respectively, and the integration over their propagation paths involves the solid angle elements $\mathrm{d}\Tilde{\Omega}_{\nu}$ and $\mathrm{d}\Tilde{\Omega}_{\overline{\nu}}$. In scenarios where the system reaches thermal equilibrium at temperature $T$, the phase-space distributions for neutrinos and antineutrinos, denoted by $\Tilde{n}(\varepsilon_{\nu})$ and $\Tilde{n}(\varepsilon_{\overline{\nu}})$, follow Fermi--Dirac statistics \cite{Salmonson:1999es}
\ie
\Tilde{n}(\varepsilon_{\nu}) = \frac{2}{h^{3}}\dfrac{1}{e^{\left({\frac{\varepsilon_{\nu}}{k \, T}}\right)} + 1}.
\fe

In this context, the energy transfer resulting from neutrino annihilation is calculated using a formulation that incorporates both Planck’s constant $h$ and Boltzmann’s constant $k$ as essential physical quantities. With these constants specified, one can derive the expression for the rate at which energy is deposited, evaluated per unit time and volume. This formulation takes into account the thermal and quantum characteristics of the process and appears in the treatment developed in \cite{Salmonson:1999es}
\ie
\frac{\mathrm{d}\mathrm{E}}{\mathrm{d}t\mathrm{d}V} = \frac{21\zeta(5)\pi^{4}}{h^{6}}\mathrm{K} \, \mathrm{G}_{f}^{2} \, \mathfrak{f}(r)(k \, T)^{9}.
\fe
The differential quantity $\mathrm{d}E/(\mathrm{d}t,\mathrm{d}V)$ plays a central role in characterizing energy transformation mechanisms near compact astrophysical objects \cite{Salmonson:1999es}. In other words, it encapsulates how local physical parameters evolve with radial distance, highlighting the significance of the temperature profile $T(r)$ in governing the thermal environment throughout the region of interest \cite{Salmonson:1999es}.

For an observer located at a radial position $r$, the local temperature measurement is influenced by the gravitational redshift, leading to the relation $T(r)\sqrt{-\mathfrak{g}_{tt}(r)} = \text{constant}$. Notice that this condition indicates the impact of spacetime curvature on thermal measurements \cite{Salmonson:1999es}. At the surface of the neutrinosphere, identified by the radial coordinate $r = R$, the temperature governing neutrino emission adheres to the constraint \cite{Salmonson:1999es}: 
\ie
T(r)\sqrt{-\mathfrak{g}_{tt}(r)} = T(R)\sqrt{-\mathfrak{g}_{tt}(R)}.
\fe  
Here, the symbol $R$ denotes the radius of the compact source responsible for the surrounding gravitational field. To simplify the computation, the temperature function $T(r)$ is rewritten based on the redshift condition introduced earlier. In this manner, taking gravitational redshift into account, one arrives at the following formulation for the neutrino luminosity \cite{Salmonson:1999es}:
\ie
\mathfrak{L}_{\infty} = -\mathfrak{g}_{tt}(R)L(R).
\fe

Additionally, the luminosity associated with a single neutrino flavor, evaluated at the radius of the neutrinosphere, can be determined using the relation below \cite{Salmonson:1999es}:
\ie
\mathfrak{L}(R) = 4 \pi R_{0}^{2}\dfrac{7}{4}\dfrac{a\,c}{4}T^{4}(R).
\fe

In this formulation, $c$ denotes the vacuum speed of light, and $a$ corresponds to the radiation constant. To account for the effects of gravitational redshift, the temperature as measured at a given radial position is adjusted accordingly. This modification yields a relation that links the local thermal state to the underlying spacetime \cite{Salmonson:1999es}:
\ie
\begin{split}
\frac{\mathrm{d}\mathrm{E}(r)}{\mathrm{d}t \, \mathrm{d}V} & = \dfrac{21\zeta(5)\pi^{4}}{h^{6}}
\mathrm{K} \, \mathrm{G}_{f}^{2} \, k^{9}\left(\dfrac{7}{4}\pi a\,c\right)^{-\frac{9}{4}}\\
& \times \mathfrak{L}_{\infty}^{\frac{9}{4}}\mathfrak{f}(r)
\left[\dfrac{\sqrt{-\mathfrak{g}_{tt}(R)}}{-\mathfrak{g}_{tt}(r)}\right]^{\frac{9}{2}} R^{-\frac{9}{2}}.
\end{split}
\fe

It is important to point out that the function $\zeta(s)$ in the expression above refers to the Riemann zeta function. For real arguments $s > 1$, this function is defined through the following infinite series:
\ie
\zeta(s) = \sum_{n=1}^{\infty} \frac{1}{n^s}.
\fe

It is worth noting that the local rate at which energy is deposited depends not only on the radial coordinate but also on the specific structure of the spacetime geometry, particularly the metric components at the surface of the compact object. To determine the total energy radiated in a curved spacetime background, one must integrate the local deposition rate over time. The calculation of the angular contribution, denoted by $\mathfrak{f}(r)$, requires a refined treatment of the variable $x$, which was previously introduced in the context of angular integration. This step proceeds as follows \cite{Salmonson:1999es,Shi:2023kid,AraujoFilho:2024mvz,Lambiase:2020iul}:
\begin{align}
x^{2}& = \sin^{2}\theta_{r}|_{\theta_{R}=0}\notag\\
&=1-\dfrac{R^{2}}{r^{2}}\dfrac{\mathfrak{g}_{tt}(r)}{\mathfrak{g}_{tt}(R)}.
\end{align}

To determine the total energy deposited in the region surrounding the compact object, one must integrate the local energy deposition density—defined per unit time and volume—across the spatial region affected by the gravitational field. This computation depends essentially on the angular contribution, which is inherently shaped by the characteristics of the spacetime geometry. The behavior of this is dictated by the form of the metric governing the background spacetime \cite{Shi:2023kid,Lambiase:2020iul,AraujoFilho:2025rzh,Shi:2025rfq}
\begin{align}
\dot{Q} & = \frac{\mathrm{d}\mathrm{E}}{\sqrt{-\mathfrak{g}_{tt}(r)}\mathrm{d}t}\\
&=\dfrac{84\zeta(5)\pi^{5}}{h^{6}}\mathrm{K}\, \mathrm{G}_{f}^{2} \, k^{9}
\left(\dfrac{7}{4}\pi a\,c\right)^{-\frac{9}{4}}
\mathfrak{L}_{\infty}^{\frac{9}{4}}\left[-\mathfrak{g}_{tt}(R)\right]^{\frac{9}{4}}\\
& \times 
R^{-\frac{3}{2}}\int_{1}^{\infty}(x-1)^4\left(x^2+4x+5\right)\sqrt{\dfrac{\mathfrak{g}_{rr}(yR)}{-\mathfrak{g}_{tt}^9(yR)}}y^2\mathrm{d}y.
\end{align}

The symbol $\dot{Q}$ denotes the total rate at which energy is deposited through the transformation of neutrino energy into electron--positron pairs at a specific radial position \cite{Salmonson:1999es}. Under conditions where this rate becomes sufficiently large, the resulting pair production may trigger significant astrophysical processes. To clarify how gravity influences this mechanism, it becomes necessary to compare the relativistic formulation of energy deposition with its Newtonian counterpart. Such a comparison reveals how curvature effects modify the efficiency of neutrino--driven energy transfer \cite{Salmonson:1999es,Lambiase:2020iul,Shi:2023kid,shi2022neutrino}
\ie
\begin{split}
\label{ratio_Q}
 \frac{\dot{Q}}{\dot{Q}_{\text{Newton}}} = 3\left[-\mathfrak{g}_{tt}(R)\right]^{\frac{9}{4}}
& \int_{1}^{\infty}(x - 1)^{4}\left(x^{2} + 4x + 5\right) \\
& \times \sqrt{\dfrac{\mathfrak{g}_{rr}(yR)}{-\mathfrak{g}_{tt}^9(yR)}}y^2\mathrm{d}y.
\end{split}
\fe
in which
\ie
\begin{split}
\mathfrak{g}_{tt}(R)&= -\left(1 - \frac{2M}{R}\right),\\
\mathfrak{g}_{tt}(yR)&= -\left(1 - \frac{2M}{yR}\right).
\end{split}
\fe
Moreover, one has as well
\begin{align}
x^{2}=1-\dfrac{1}{y^{2}}\frac{1-\dfrac{2M}{yR}}{1-\dfrac{2M}{R}}.
\end{align}

\begin{figure*}
\centering
\includegraphics[height=6.5cm]{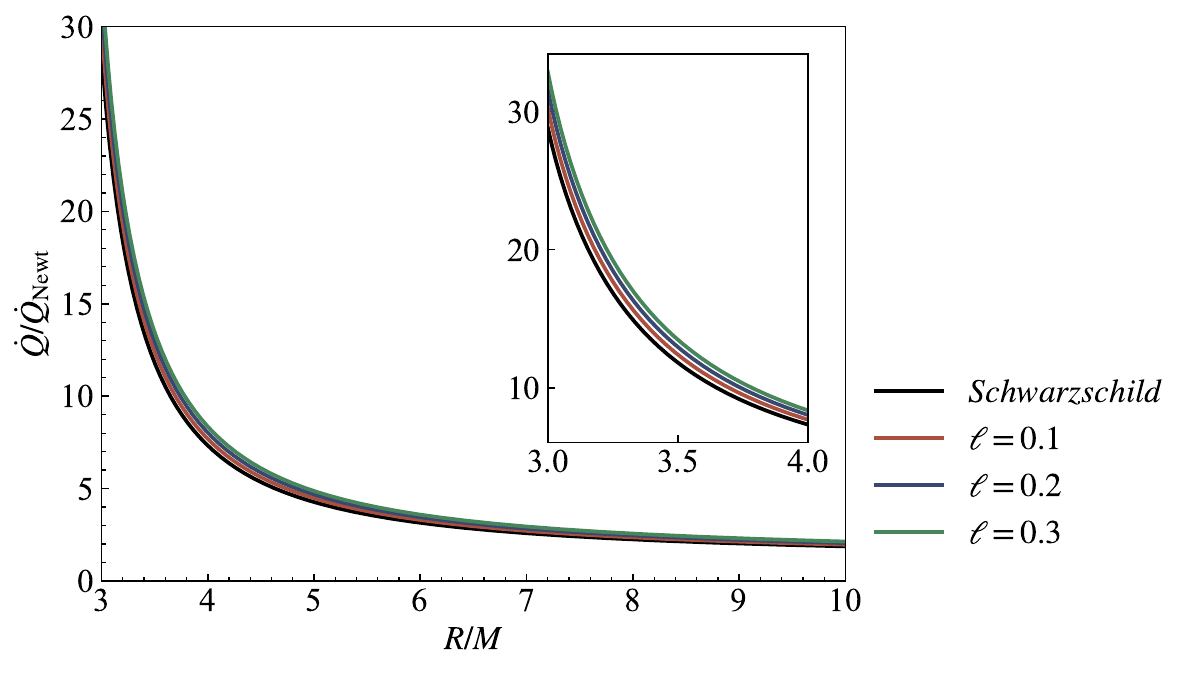}
\caption{The plot displays the behavior of the ratio $\dot{Q}/\dot{Q}_{\text{Newton}}$ as a function of $R/M$, considering different values of the Lorentz--violating parameter $\ell$.}
\label{esndedaradgydadeapodsaiidatoadna}
\end{figure*}

\begin{figure*}
\centering
\includegraphics[height=6.5cm]{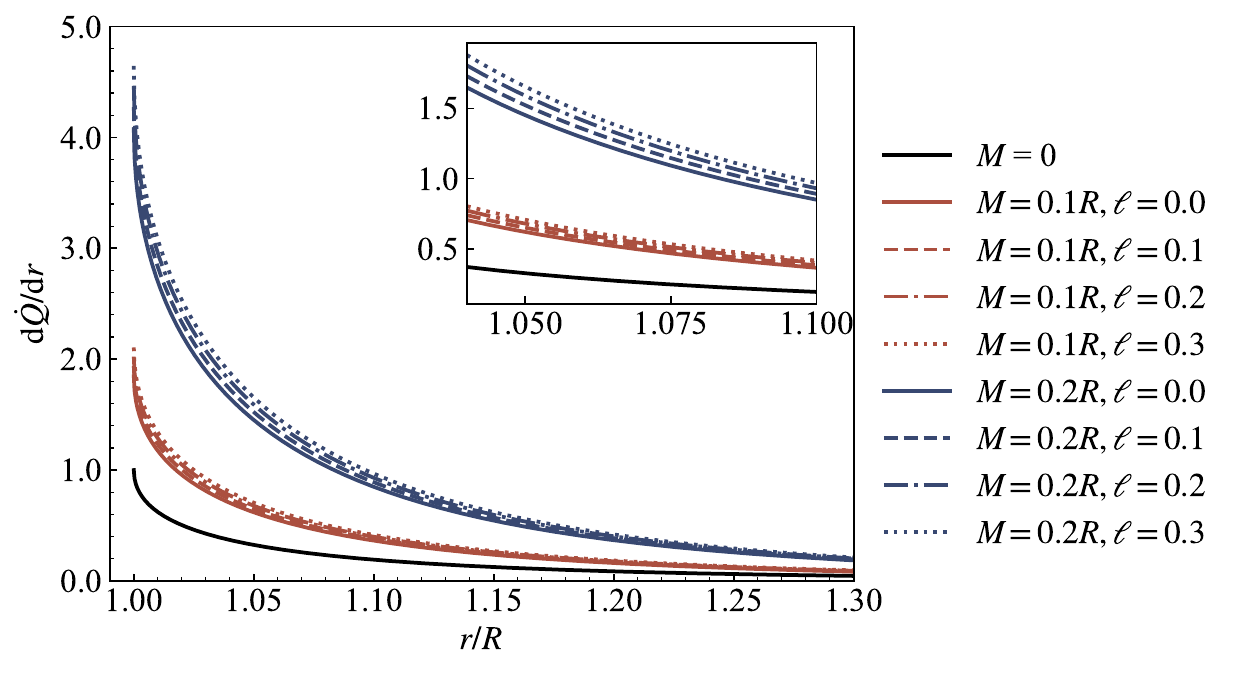}
\caption{The radial profile of the differential energy deposition rate, $\mathrm{d}\dot{Q}/\mathrm{d}r$, is analyzed for different values of the compactness ratio $M/R$. In the Newtonian regime—corresponding to the case where $M = 0$—the relation reduces to a simpler form, resulting in $\mathrm{d}\dot{Q}/\mathrm{d}r = 1$ precisely at the boundary $r = R$.}
\label{dQvfgdr}
\end{figure*}
 
\begin{table}[h!]
\centering
\caption{The emission rate $\dot{Q}$, expressed in erg/s, is computed for a range of values of $\ell$ as well as different choices of the dimensionless compactness ratio $R/M$.}
\label{taabb::dootQ}
\begin{tabular}{ccc}
\hline\hline
$\ell$ & $R/M$ & $\dot{Q}$ \((\text{erg/s})\) \\
\hline
0.0 & 0 & $1.50 \times 10^{50}$ \\
\hline
\multirow{2}{*}{0.0} & 3 & $4.32 \times 10^{51}$ \\
                      & 4 & $1.10 \times 10^{51}$ \\
\hline
\multirow{2}{*}{0.1} & 3 & $4.53 \times 10^{51}$ \\
                      & 4 & $1.15 \times 10^{51}$ \\
\hline
\multirow{2}{*}{0.2} & 3 & $4.73 \times 10^{51}$ \\
                      & 4 & $1.20 \times 10^{51}$ \\
\hline
\multirow{2}{*}{0.3} & 3 & $4.93 \times 10^{51}$ \\
                      & 4 & $1.25 \times 10^{51}$ \\
\hline\hline
\end{tabular}
\end{table}

Clearly, the integrand in Eq.~\eqref{ratio_Q} is smooth and non-oscillatory, and the integration interval $[1,\infty]$ is free of discontinuities. The endpoints $y\to1$ and $y\to\infty$ have limit values of $5R^5\sqrt{1+\ell}/(R-2)^5$ and $0$, respectively. The Simpson's integration technique may be used to evaluate such an integrand, and the Python utility \textit{scipy.integrate} can be used to acquire the numerical integration result. We have set the relative and absolute error tolerances to $1.49\times10^{-8}$. It is predicted that the absolute error of the outcomes throughout the numerical computing process ranges from $3.8\times10^{-11}$ to $4.8\times10^{-11}$. To investigate the impact of Lorentz symmetry breaking on neutrino--antineutrino annihilation, we reconstruct the formalism for energy deposition using a Schwarzschild--like solution modified within the context of Bumblebee gravity. This approach allows us to analyze deviations from general relativity induced by the parameter $\ell$. Fig.~\ref{esndedaradgydadeapodsaiidatoadna} presents the results for the energy conversion efficiency ratio, given by Eq. (\ref{ratio_Q}), for a range of $\ell$ values. The comparison between the standard relativistic case and the Lorentz--violating extensions reveals an unexpected enhancement in the energy deposition rate across all curves, despite the parameter $\ell$ varying from $0$ to $0.3$. Notably, these enhancements preserve similar radial behavior when compared to general relativity. This result contrasts with the suppression patterns found in gravitational modifications arising from the Kalb--Ramond gravity recently addressed in the literature \cite{Shi:2025rfq}.

The data presented in Tab.~\ref{taabb::dootQ} reveal that, assuming a typical neutrino luminosity on the order of $10^{53}$ erg/s measured at spatial infinity, along with a fixed emission radius of $R = 20\,\mathrm{km}$ and parameter $D = 1.23$, the presence of Lorentz--violating effects in the Bumblebee framework leads to a slight increase in the overall energy deposition rate. This outcome stands in contrast to the suppression observed in the Kalb--Ramond black hole scenario \cite{Shi:2025rfq}, where the altered geometry results in reduced energy transfer.

The radial profiles of the differential energy deposition rate, $\mathrm{d}\dot{Q}/\mathrm{d}r$, depicted in Fig.~\ref{dQvfgdr}, indicate a sharper gradient in the Bumblebee gravity scenario compared to the Kalb--Ramond framework. This distinction arises from the differing ways each model alters the structure of null geodesics. In the Bumblebee case, the parameter $\ell$ modifies the asymptotic form of the metric component $\mathfrak{g}_{rr}$, which effectively enhances the radial variation of the neutrino--antineutrino collision cross--section. On the other hand, the Kalb--Ramond modification introduces additional curvature terms that increase the divergence of neutrino trajectories, thereby reducing the spatial concentration of interaction events and leading to a shallower energy deposition gradient \cite{Shi:2025rfq}.

A comparison between the two gravitational models uncovers contrasting mechanisms that shape energy deposition behavior. In other words, in the Bumblebee scenario, the Lorentz--violating parameter $\ell$ introduces modifications through the coupling of the background vector field. These alterations partially mitigate energy loss in regions dominated by intense gravitational fields. In contrast, the Kalb--Ramond model amplifies the suppression of energy conversion by deepening the gravitational potential via tensor field contributions, thereby reducing the likelihood of efficient neutrino--antineutrino interactions. The results point to the possibility that Lorentz--violating effects in the Bumblebee model could open up a new pathway for energy production in extremely dense astrophysical systems, like the remnants left behind after neutron star mergers. This mechanism may help explain sudden spikes in brightness often seen in short--duration gamma--ray bursts \cite{Poddar:2022svz}.


\section{Neutrino phase evolution and conversion probability}

To investigate the trajectory of neutrinos linked to the $k$--th eigenmode in a spacetime exhibiting spherical symmetry, one may adopt the Lagrangian methodology presented in Ref.~\cite{neu18}. This framework, applied to the metric structure introduced in Eq.~(\ref{mmainin}), enables a systematic derivation of the governing equations for neutrino motion by exploiting variational principles adapted to curved backgrounds as follows
\begin{align}
\mathcal{L}
& = -\frac{1}{2}  m_{k} \mathfrak{g}_{tt}(r)\left(\frac{\mathrm{d}t}{\mathrm{d}\tau}\right)^2-\frac{1}{2}m_{k}\mathfrak{g}_{rr}(r)\left(\frac{\mathrm{d}r}{\mathrm{d}\tau}\right)^2 \notag\\
& \quad -\dfrac{1}{2}m_{k}r^2\left(\frac{\mathrm{d}\theta}{\mathrm{d}\tau}\right)^2  
 -\frac{1}{2}m_{k}r^2\sin^2\theta\left(\frac{\mathrm{d}\varphi}{\mathrm{d}\tau}\right)^2.
\end{align}

By confining the motion to the equatorial plane, where $\theta = \pi/2$, the resulting dynamics involve only a limited set of non--zero momentum components. To obtain these, one starts by introducing the canonical momentum $p_{\mu}$ through the relation $p_{\mu} = \partial \mathcal{L} / \partial (\mathrm{d}x^{\mu} / \mathrm{d}\tau)$, where $\mathcal{L}$ denotes the Lagrangian and $\tau$ represents the proper time parameter. The quantity $m_k$ designates the mass associated with the $k$--th neutrino eigenstate. Detailed forms of the relevant momentum components that survive under this symmetry reduction are provided in Refs.~\cite{neu60,Shi:2024flw}
\begin{align}
p^{(k)t} &= -m_{k}\mathfrak{g}_{tt}(r)\frac{\mathrm{d}t}{\mathrm{d}\tau} = -E_{k}, \\
p^{(k)r} &= m_{k}\mathfrak{g}_{rr}(r)\frac{\mathrm{d}r}{\mathrm{d}\tau}, \\
p^{(k)\varphi} &= m_{k}r^2\frac{\mathrm{d}\varphi}{\mathrm{d}\tau} = J_{k}.
\end{align}

Notice that, as one should expect, the relativistic dynamics of a neutrino belonging to the $k$--th mass eigenstate are governed by the mass--shell relation, which establishes the constraint $\mathfrak{g}^{\mu\nu} p_\mu p_\nu = -m_k^2$. This condition guarantees compatibility between the particle's motion and the underlying spacetime geometry, ensuring that its trajectory remains consistent with general relativistic principles \cite{neu54}
\begin{align}
-m_{k}^2 =\mathfrak{g}^{tt}p_t^2+\mathfrak{g}^{rr}p_r^2+\mathfrak{g}^{\varphi\varphi}p_{\varphi}^2.
\end{align}

Furthermore, in areas where the influence of gravity is minimal, neutrino oscillation studies typically rely on the plane wave framework \cite{neu54,neu53}. Since weak interactions dictate both production and detection mechanisms, neutrinos are not detected in mass eigenstates but rather in superpositions known as flavor eigenstates, as examined in detail in previous works \cite{neu62,neu61,neu63,Shi:2024flw}
\begin{align}
\ket{\nu_{\alpha}} = \sum U_{\alpha i}^{*}\ket{\nu_{i}}.
\end{align}

The evolution of neutrinos through spacetime is most effectively characterized using mass eigenstates, labeled as $\ket{\nu_i}$, each following a unique trajectory influenced by its mass. Despite this, neutrinos are produced and measured in flavor states—$\nu_e$, $\nu_\mu$, and $\nu_\tau$—indexed by $\alpha = e, \mu, \tau$, due to the nature of weak interactions. The connection between these physical flavor states and the underlying mass basis is governed by a unitary transformation involving a $3 \times 3$ mixing matrix $U$, as outlined in Ref.~\cite{neu41}. To model their propagation, one typically associates spacetime coordinates $\left(t_S, \bm{x}_S\right)$ and $\left(t_D, \bm{x}_D\right)$ with the emission and detection points, respectively. Each mass eigenstate wave packet then evolves along the spacetime trajectory linking these two locations
\begin{align}
\ket{\nu_{i}\left(t_{D},\bm{x}_{D}\right)} = \exp({-\mathbbm{i}\Tilde{\Phi}_{i}})\ket{\nu_{i}\left(t_{S},\bm{x}_{S}\right)}.
\end{align}
As a result, each mass eigenstate accumulates a distinct phase while traveling between the source and detector, which can be determined through the following relation:
\begin{align}
\Tilde{\Phi}_{i}=\int_{\left(t_{S},\bm{x}_{S}\right)}^{\left(t_{D},\bm{x}_{D}\right)}\mathfrak{g}_{\mu\nu}p^{(i)\mu}\mathrm{d}x^{\nu}.
\end{align}

This framework investigates neutrino flavor transitions by following the particle’s propagation from the emission site to the detection point. During this journey, a neutrino initially in the flavor state $\nu_{\alpha}$ can convert into a different flavor $\nu_{\beta}$. The corresponding probability of this flavor conversion is determined by the expression given below:
\begin{align}
\mathcal{P}_{\alpha\beta}
& = |\left\langle \nu_{\beta}|\nu_{\alpha}\left(t_{D}, \bm{x}_{D}\right)\right\rangle|^2 \\
& = \sum_{i,j} U_{\beta i}U_{\beta j}^{*} U_{\alpha j} U_{\alpha i}^{*}\,  \exp{[-\mathbbm{i}(\Tilde{\Phi}_{i}-\Tilde{\Phi}_{j})]}.
\end{align}

In the context of a bumblebee black hole spacetime, this investigation considers neutrinos whose trajectories lie entirely in the equatorial plane, defined by $\theta = \pi/2$. Under such conditions, the phase accumulated by each neutrino along its path is described by the expression presented below:
\begin{align}
\label{Pgefhi}
\Tilde{\Phi}_{k} & = \int_{\left(t_{S},\bm{x}_{S}\right)}^{\left(t_{D}, \bm{x}_{D}\right)} \mathfrak{g}_{\mu\nu} p^{(k)\mu}\mathrm{d}x^{\nu}\notag\\
& = \int_{\left(t_{S},\bm{x}_{S}\right)}^{\left(t_{D}, \bm{x}_{D}\right)}\left[E_{k}\mathrm{d}t - p^{(k)r}\mathrm{d}r-J_{k}\mathrm{d}\varphi\right] \notag\\
& = \pm\frac{m_{k}^2}{2E_0}\int_{r_{S}}^{r_{D}}\sqrt{-\mathfrak{g}_{tt}\mathfrak{g}_{rr}}\left(1-\dfrac{b^2|\mathfrak{g}_{tt}|}{\mathfrak{g}_{\varphi\varphi}}\right)^{-\frac{1}{2}}\mathrm{d}r.
\end{align}

When the gravitational field is sufficiently weak, satisfying the condition $M/r \ll 1$, the integrand appearing in Eq.~(\ref{Pgefhi}) can be expanded as a series as given below
\begin{align}
&\quad\sqrt{-\mathfrak{g}_{tt}\mathfrak{g}_{rr}}\left(1-\dfrac{b^2|\mathfrak{g}_{tt}|}{\mathfrak{g}_{\varphi\varphi}}\right)^{-\frac{1}{2}}\notag\\
&\simeq\left(1+\dfrac{\ell}{2}\right)\left[\dfrac{r}{\sqrt{r^2-b^2}}-\dfrac{b^2 M}{\left(r^2-b^2\right)^{\frac{3}{2}}}\right].
\end{align}
Consequently, the resulting phase takes the form given by the following expression:
\begin{align}
\Tilde{\Phi}_k&=\dfrac{m_k^2}{2E_0}\left(1+\dfrac{\ell}{2}\right)\Biggl[\sqrt{r_D^2-b^2}-\sqrt{r_S^2-b^2}\notag\\
&\quad+M\left(\dfrac{r_D}{\sqrt{r_D^2-b^2}}-\dfrac{r_S}{\sqrt{r_S^2-b^2}}\right)\Biggr].
\end{align}

Within this approach, the average energy of relativistic neutrinos originating from the source is given by $E_0 = \sqrt{E_k^2 - m_k^2}$, where $E_k$ denotes the energy and $m_k$ the mass of the $k$--th mass eigenstate. The quantity $b$, known as the impact parameter, is addressed in Ref.~\cite{neu18}. As the neutrinos move through the curved background, they approach the central object down to a minimum radial coordinate $r = r_0$. In the regime of a weak gravitational field, the equation determining this point of closest approach can be solved analytically using appropriate approximations
\begin{align}
\left(\dfrac{\mathrm{d}r}{\mathrm{d}\varphi}\right)_0=\pm\dfrac{\mathfrak{g}_{\varphi\varphi}}{b^2}\sqrt{\dfrac{1}{-\mathfrak{g}_{tt}\mathfrak{g}_{rr}}-\dfrac{b^2}{\mathfrak{g}_{rr}\mathfrak{g}_{\varphi\varphi}}}=0.
\end{align}

The minimum radial distance $r_0$ can be determined by analyzing its orbital dynamics under the influence of a weak gravitational field. This value emerges as a solution to the equation that characterizes the particle’s trajectory in such a regime
\begin{align}
\label{r0}
r_0 \simeq b-M.
\end{align}

To compute the full phase accumulated by a neutrino as it travels from the emission point, passes through its point of nearest approach, and reaches the detector, one combines the expression in Eq.~(\ref{Pgefhi}) with the approximation for $r_0$ given in Eq.~(\ref{r0}). This approach yields the following result:
\begin{align}
\label{pphhiii}
&\quad\Tilde{\Phi}_{k}\left(r_{S}\to r_{0} \to r_{D}\right)\notag\\
&\simeq \frac{{m}_{k}^2}{2E_0}\left(1+\dfrac{\ell}{2}\right)
\Biggl[\sqrt{r_D^2-r_0^2}+\sqrt{r_S^2-r_0^2}\notag\\
&\quad+M\left(\sqrt{\dfrac{r_D-r_0}{r_D+r_0}}+\sqrt{\dfrac{r_S-r_0}{r_S+r_0}}\right)\Biggr],
\end{align}
so that
\begin{align}
\Tilde{\Phi}_{k}
& \simeq \frac{{m}_{k}^2}{2E_0}\left(1+\dfrac{\ell}{2}\right)
\Biggl[\sqrt{r_D^2-b^2}+\sqrt{r_S^2-b^2}\notag\\
& \quad+M\Biggl(\dfrac{b}{\sqrt{r_D^2-b^2}}+\dfrac{b}{\sqrt{r_S^2-b^2}}\notag\\
&\quad+\sqrt{\dfrac{r_D-b}{r_D+b}}+\sqrt{\dfrac{r_S-b}{r_S+b}}\Biggr)\Biggr].
\end{align}

The subsequent step involves expanding the previously obtained expression in a power series of $b/r_{S,D}$, assuming that the impact parameter is much smaller than both the source and detector distances, i.e., $b \ll r_{S,D}$. Keeping terms up to second order, $\mathcal{O}\left(b^2/r_{S,D}^2\right)$, one arrives at the following approximation:
\begin{align}
\label{Phi_k}
\Tilde{\Phi}_k=\dfrac{m_k^2}{2E_0}\left(1+\dfrac{\ell}{2}\right)(r_D+r_S)\left(1-\dfrac{b^2}{2r_Dr_S}+\dfrac{2M}{r_D+r_S}\right).
\end{align}

As the Lorentz--violating parameter $\ell$ increases, one observes a clear increment in the phase acquired by neutrinos during their journey. For the purposes of this evaluation, the parameters are set to $E_0 = 10\,\mathrm{MeV}$, $r_D = 10\,\mathrm{km}$, and $r_S = 10^5 r_D$.

In this context, the curvature of spacetime induces gravitational lensing, which deflects the neutrino trajectories. To analyze the probability of flavor transitions in the vicinity of the black hole, it becomes necessary to evaluate the phase shifts arising from the various possible propagation paths \cite{Shi:2024flw,AraujoFilho:2025rzh}
\begin{align}
\Delta\Tilde{\Phi}_{ij}^{pq}
&= \Tilde{\Phi}_i^{p}-\Tilde{\Phi}_j^{q}\notag\\
&= \Delta m_{ij}^2 A_{pq}+\Delta b_{pq}^2 B_{ij},
\end{align}
where
\begin{align}
\label{Delta_m}
\Delta m_{ij}^2 & = m_i^2 - m_j^2,\\
\Delta b_{pq}^2 & = b_{p}^2-b_{q}^2,\\
A_{pq} & = \frac{r_{S} + r_{D}}{2 E_0}\left(1+\dfrac{\ell}{2}\right)\left[1+\dfrac{2M}{r_D+r_S}-\dfrac{\sum b_{pq}^2}{4r_Dr_S}\right],\\
B_{ij} & = -\frac{\sum m_{ij}^2}{8E_0}\left(1+\dfrac{\ell}{2}\right)\left(\frac{1}{r_{D}} + \frac{1}{r_{S}}\right),\\
\sum b_{pq}^2 & = b_{p}^2 + b_{q}^2,\\
\label{sum_m}
\sum m_{ij}^2 & = m_i^2 + m_j^2.
\end{align}

To distinguish between the phases associated with various neutrino paths, a superscript notation is adopted—$\Tilde{\Phi}_{i}^{p}$—where each label $p$ denotes a distinct trajectory characterized by its corresponding impact parameter $b_p$. In our case, the resulting phase differences responsible for oscillation phenomena depend on the masses $m_i$ of the neutrino eigenstates, the mass--squared differences $\Delta m_{ij}^2$, and the geometric features of the spacetime. Notably, in the limit $\ell \to 0$, the expression for the phase shift converges to the standard result reported in Ref.~\cite{neu53}.

Additionally, it is worth emphasizing that while the coefficient $B_{ij}$ encapsulates all mass--related contributions, the Lorentz--violating term $\ell$ manifests exclusively through modifications in the $A_{pq}$ factor, which influences both the phase magnitude and the amplitude of oscillations.


\section{Gravitational deflection of neutrino trajectories}

\begin{figure}
    \centering
    \includegraphics[scale=0.45]{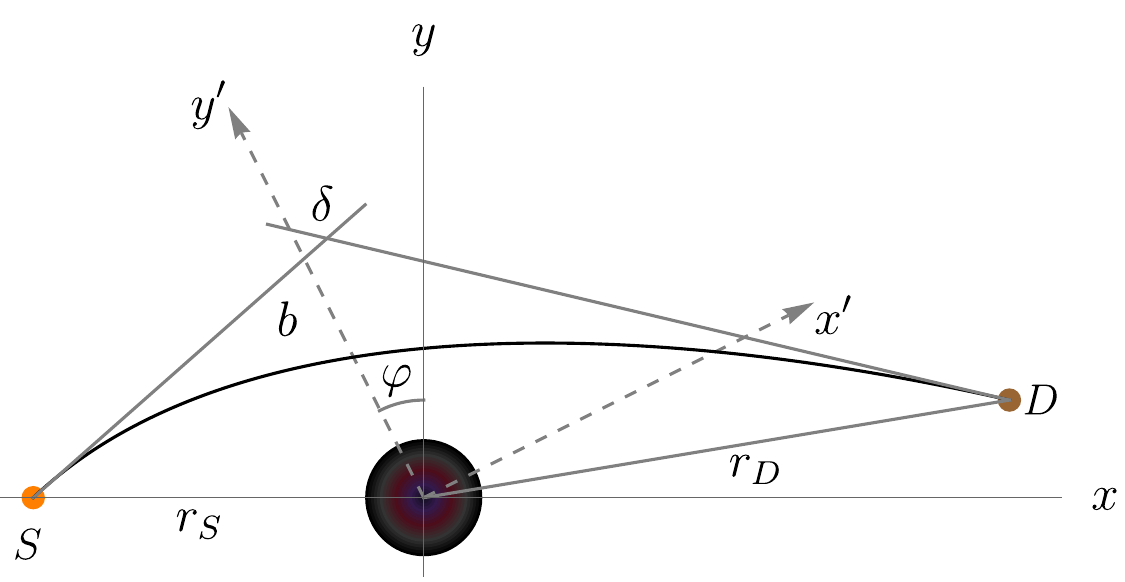}
    \caption{The illustration demonstrates the influence of weak gravitational lensing on neutrino trajectories within a curved spacetime background. The location denoted by $S$ represents the source of emission, while the point marked $D$ corresponds to the position of the detector.}
    \label{laesnsssisnsg}
\end{figure}

In the presence of a strong gravitational field generated by a compact and massive object, neutrino trajectories can be significantly bent, causing deviations from purely radial motion due to lensing phenomena \cite{neu54}. This bending enables multiple paths to connect a common source with the same detection point $D$, as illustrated in Fig.~\ref{laesnsssisnsg}. Under such circumstances, the conventional description of a neutrino flavor state must be extended to incorporate a coherent superposition over all contributing trajectories \cite{neu64,neu62,neu65,Shi:2024flw,neu56,neu63,AraujoFilho:2025rzh}:
\begin{align}
|\nu_{\alpha}(t_{D},x_{D})\rangle = N\sum_{i}U_{\alpha i}^{\ast}
\sum_{p} e^{- \mathbbm{i} \Tilde{\Phi}_{i}^{p}}|\nu_{i}(t_{S}, x_{S})\rangle.
\end{align}

In this process, every possible neutrino trajectory is labeled by an index $p$. Given that all such paths converge at the same detection location, the transition from an initial flavor state $\nu_{\alpha}$ to a final state $\nu_{\beta}$ arises from the interference among the amplitudes corresponding to each distinct route. The resulting transition probability is obtained by coherently summing over all contributing paths, and is expressed as shown below \cite{Shi:2024flw,neu56,neu62,neu63,neu64,neu65}:
\begin{align}
\label{nasndkas}
\mathcal{P}_{\alpha\beta}^{\mathrm{lensing}} & = |\langle \nu_{\beta}|\nu_{\alpha}(t_{D}, x_{D})\rangle|^{2}\notag\\
& =|N|^{2}\sum_{i, j}U_{\beta i}U_{\beta j}^{\ast}U_{\alpha j}U_{\alpha j}^{\ast}\sum_{p, q}e^{\Delta\Tilde{\Phi}_{ij}^{pq}}.
\end{align}
Based on this formulation, the normalization constant is exhibited below
\begin{align}
|N|^{2} = \left[\sum_{i}|U_{\alpha i}|^{2}\sum_{p,q}\exp\left(-\mathbbm{i}\Delta\Tilde{\Phi}_{ij}^{pq}\right)\right]^{-1}.
\end{align}

The previously defined phase difference $\Delta\Tilde{\Phi}_{ij}^{pq}$ plays a fundamental role in evaluating the neutrino oscillation probability in scenarios influenced by gravitational lensing. As we mentioned before, this probability depends on several factors, including the individual neutrino masses, the differences in their squared masses, and the properties of the curved geometry induced by the black hole, as expressed in Eq.~(\ref{nasndkas}). The resulting formalism shares similarities with those derived in spherically symmetric backgrounds such as the Schwarzschild spacetime \cite{neu53,Shi:2024flw,AraujoFilho:2025rzh}.

Now, our analysis turns to the effect of gravitational lensing on neutrino flavor conversion, emphasizing the influence due to $\ell$. Within this lensing configuration, considering a two--flavor neutrino system, the probability of transition from $\nu_{\alpha}$ to $\nu_{\beta}$ is determined under the weak--field approximation, taking into account the spatial arrangement of the emission point, the detector and lensing region \cite{neu53,neu54,Shi:2024flw,neu65}
\begin{align}
\label{asdPdadbd2}
\mathcal{P}_{\alpha\beta}^{\mathrm{lensing}}
&=\left|N\right|^2\biggl\{2\sum_i\left|U_{\beta i}\right|^2\left|U_{\alpha i}\right|^2\left[1+\cos\left(\Delta b_{12}^2B_{ii}\right)\right]\notag\\
&\quad+\sum_{i\neq j}U_{\beta i}U_{\beta j}^*U_{\alpha j}U_{\alpha i}^*\notag\\
&\quad\times\left[\exp\left(-\mathbbm{i}\Delta m_{ij}^2 A_{11}\right)+\exp\left(-\mathbbm{i}\Delta m_{ij}^2 A_{22}\right)\right]\notag\\
&\quad+\sum_{i\neq j}U_{\beta i}U_{\beta j}^*U_{\alpha j}U_{\alpha i}^*\notag\\
&\quad\times\exp\left(-\mathbbm{i}\Delta b_{12}^2B_{ij}\right)\exp\left(-\mathbbm{i}\Delta m_{ij}^2A_{12}\right)\notag\\
&\quad+\sum_{i\neq j}U_{\beta i}U_{\beta j}^*U_{\alpha j}U_{\alpha i}^*\notag\\
&\quad\times\exp\left(\mathbbm{i}\Delta b_{21}^2B_{ij}\right)\exp\left(-\mathbbm{i}\Delta m_{ij}^2A_{21}\right)\biggr\}.
\end{align}

It is worth mentioning that Eq.~(\ref{asdPdadbd2}) formulates the probability of neutrino flavor conversion as a structured sum of terms, each enclosed in braces and classified by the specific pairings of mass eigenstates and the propagating paths. The component with $i = j$ represents the independent evolution of a single mass eigenstate, where no interference occurs. In contrast, when $i \neq j$ and $p = q$, the term corresponds to quantum interference between distinct mass eigenstates traveling along the same geodesic, each acquiring a different phase.

Further interference effects arise in the mixed case where both the mass indices and the trajectory labels are different ($i \neq j$, $p \neq q$). These cross terms are evaluated separately for $p < q$ and $p > q$, reflecting the inherent asymmetry in phase accumulation due to variations in the path lengths and the background curvature.

When the system is limited to two neutrino flavors, the mathematical structure of oscillations becomes more tractable. The relation between flavor and mass eigenstates is then governed by a $2 \times 2$ unitary matrix that depends exclusively on a single parameter: the mixing angle $\alpha$ \cite{neu43}
\begin{align}
\label{U}
U\equiv\left(\begin{matrix}
\cos\alpha&\sin\alpha\\
-\sin\alpha&\cos\alpha
\end{matrix}\right).
\end{align}

Substituting the mixing matrix from Eq.~(\ref{U}) into the general transition probability framework given in Eq.~(\ref{asdPdadbd2}) leads to a concrete expression for the probability associated with the flavor transition
$\nu_e \rightarrow \nu_\mu$
\begin{align}
\label{proballl}
\mathcal{P}_{\alpha\beta}^{\mathrm{lensing}}
&=\left|N\right|^2\sin^{2}2\alpha\notag\\
&\quad\times\biggl[\sin^2\left(\dfrac{1}{2}\Delta m_{12}^2A_{11}\right)+\sin^2\left(\dfrac{1}{2}\Delta m_{12}^2A_{22}\right)\notag\\
&\quad+\dfrac{1}{2}\cos\left(\Delta b_{12}^2B_{11}\right)+\dfrac{1}{2}\cos\left(\Delta b_{12}^2B_{22}\right)\notag\\
&\quad-\cos\left(\Delta b_{12}^2B_{12}\right)\cos\left(\Delta m_{12}^2A_{12}\right)\biggr].
\end{align}

Considering the leptonic mixing matrix defined in Eq.\~(\ref{U}) along with the phase differences accumulated along each distinct neutrino trajectory, the normalization constant takes the following form:
\begin{align}
\left|N\right|^2&=\biggl[2+2\cos^2\alpha\cos\left(\Delta b_{12}^2B_{11}\right)\notag\\
&\quad+2\sin^2\alpha\cos\left(\Delta b_{12}^2B_{22}\right)\biggr]^{-1}.
\end{align}


\section{Numerical investigation}

To analyze the behavior of neutrino oscillations in the curved spacetime generated by the black hole under consideration, one must compute the lensing probabilities outlined in Eq.~(\ref{proballl}). In the adopted coordinate system $(x, y)$, the gravitational lens is located at the origin, with the source and detector placed at radial distances $r_S$ and $r_D$, respectively. For computational convenience, a new coordinate frame $(x', y')$ is defined by rotating the original system through an angle $\varphi$. This coordinate transformation leads to the following relation \cite{neu53,Shi:2024flw}: 
\begin{align}
x' = x\cos\varphi + y\sin\varphi, \quad y' = -x\sin\varphi + y\cos\varphi .
\nonumber
\end{align}

It is important to mention that, by setting $\varphi = 0$, results in a geometric configuration where the source, lens, and detector lie colinearly within the plane, effectively aligning all three along a single straight path. This simplification places each element along the same axis, streamlining the analysis of neutrino propagation.

According to the results presented in Refs.~\cite{neu53,Shi:2024flw}, the curvature--induced deviation in a neutrino’s path—quantified by the deflection angle $\delta$—is intrinsically linked to the impact parameter $b$. Source $S$ produces neutrinos, which are then gravitationally lensed by the bumblebee black hole and finally detected at D. In the $(x,y)$ coordinate system, the physical distances from the source to the lens and the lens to the detector are denoted by the variables $r_S$ and $r_D$, respectively. In Ref~\cite{14}, the equation for the gravitational lensing of the bumblebee black hole has been derived carefully and strictly. This relationship is expressed through the following equation \cite{14}:
\begin{align}
\label{delta}
\delta \sim\frac{y_{D}'- b}{x_{D}'}=-\dfrac{\pi\ell}{2}-\dfrac{4M}{b}.
\end{align}

By positioning the detector at $(x_{D}', y_{D}')$ in the rotated coordinate system and employing the identity $\sin\varphi = b/r_S$, the deflection angle described in Eq.~(\ref{delta}) can be reformulated as follows:
\begin{align}
\label{solve_b}
&\quad\left(4Mx_D+by_D+\dfrac{\pi\ell}{2}bx_D\right)\sqrt{1-\dfrac{b^2}{r_S^2}}\notag\\
&=b^2\left(\dfrac{x_D}{r_S}+1-\dfrac{\pi\ell y_D}{2r_S}\right)-\dfrac{4Mby_D}{r_S}.
\end{align}

We computed the ratio of $b_{1,2}^2/r_{S,D}^2$ by doing numerical calculations for $b$ in Eq.~\eqref{solve_b}, yielding the values displayed in Tab.~\ref{tab:ratio}. Four distinct angles $\phi=0,0.001,0.002,0.003$ were chosen as samples. We set the length to $\mathbf{fm}$ to correspond to the unit $\mathbf{GeV}$. The value of $b_{1,2}^2$ is far lower than that of $r_{S,D}^2$, as Tab.~\ref{tab:ratio} makes evident. The correctness of employing the expansion $\mathcal{O}\left(b^2/r_{S,D}^2\right)$ in Eq.~\eqref{Phi_k} is further demonstrated by the numerical solution of deflection angle Eq.~\eqref{delta}.

\begin{table}[!ht]
\centering
\caption{The ratio about $b_{1,2}^2/r_{S,D}^2$ with $r_D=10^{26}\,\mathrm{fm}$, $r_S=10^{31}\,\mathrm{fm}$ and $\ell=1\times10^{-10}$.}
\begin{tabular}{|c|c|c|c|}
\hline
              & $\phi=0$                & $\phi=0.0015$           & $\phi=0.003$            \\ \hline
$b_1^2/r_D^2$ & $5.99994\times10^{-8}$  & $2.36843\times10^{-6}$  & $9.11939\times10^{-6}$  \\ \hline
$b_1^2/r_S^2$ & $5.99994\times10^{-18}$ & $2.36843\times10^{-16}$ & $9.11939\times10^{-16}$ \\ \hline
$b_2^2/r_D^2$ & $5.99994\times10^{-8}$  & $1.51996\times10^{-6}$  & $3.94751\times10^{-6}$  \\ \hline
$b_2^2/r_S^2$ & $5.99994\times10^{-18}$ & $1.51996\times10^{-16}$ & $3.94751\times10^{-16}$ \\ \hline
\end{tabular}
\label{tab:ratio}
\end{table}

\begin{figure*}
\centering
\includegraphics[height=5cm]{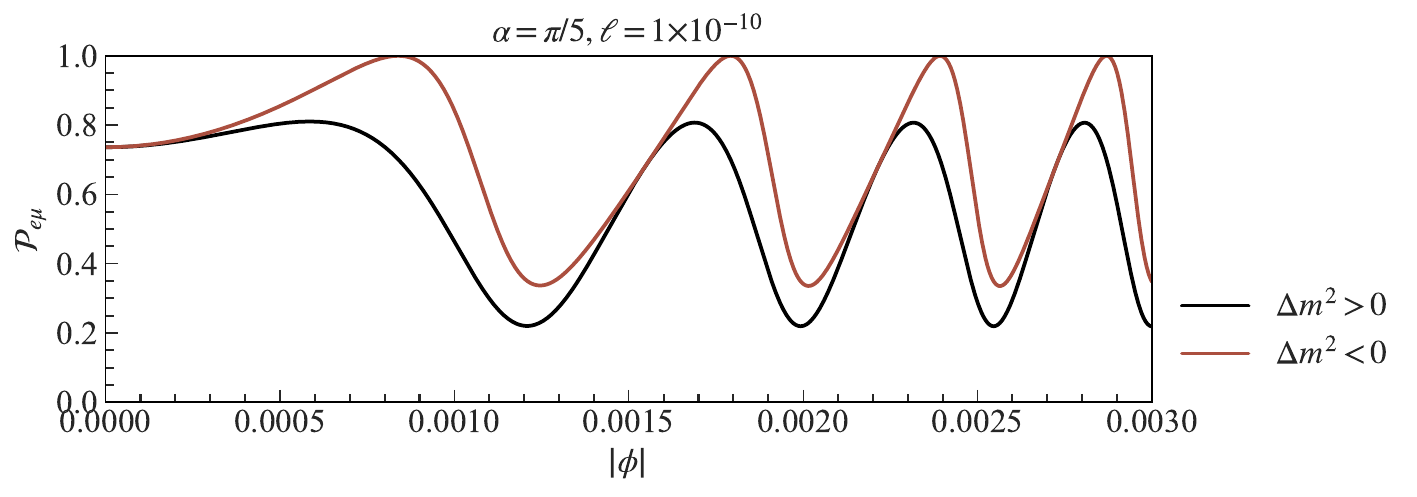}
\includegraphics[height=5cm]{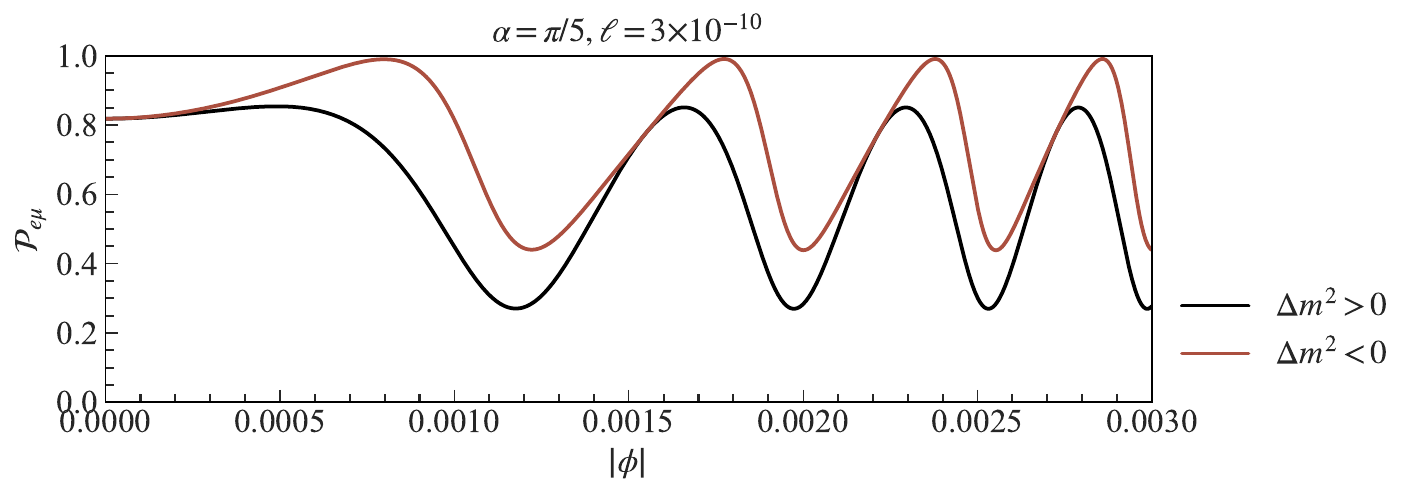}
\includegraphics[height=5cm]{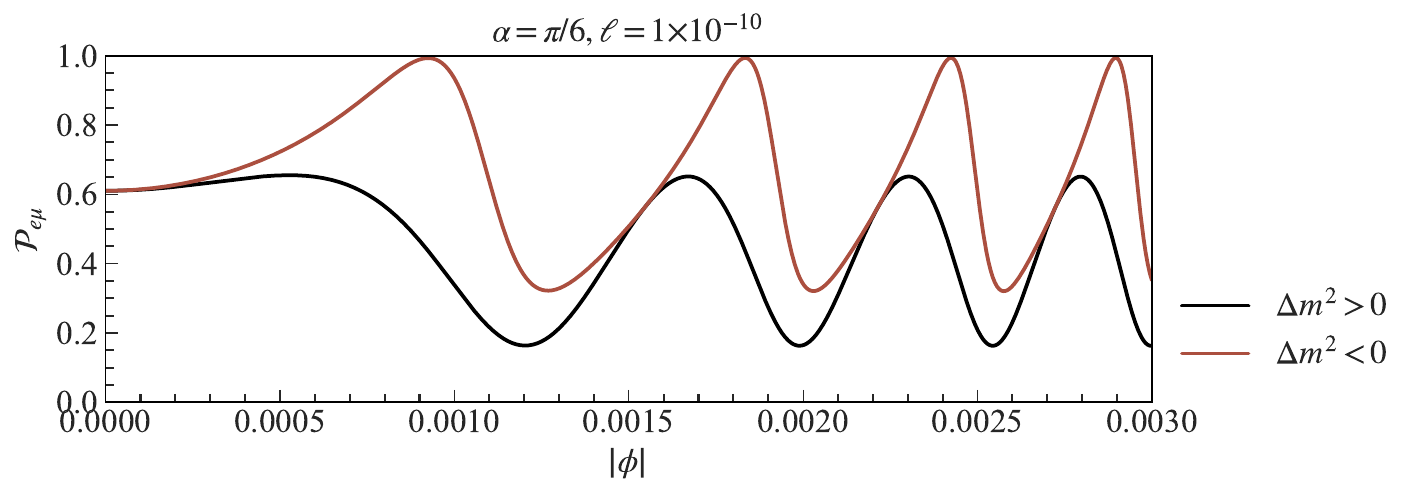}
\includegraphics[height=5cm]{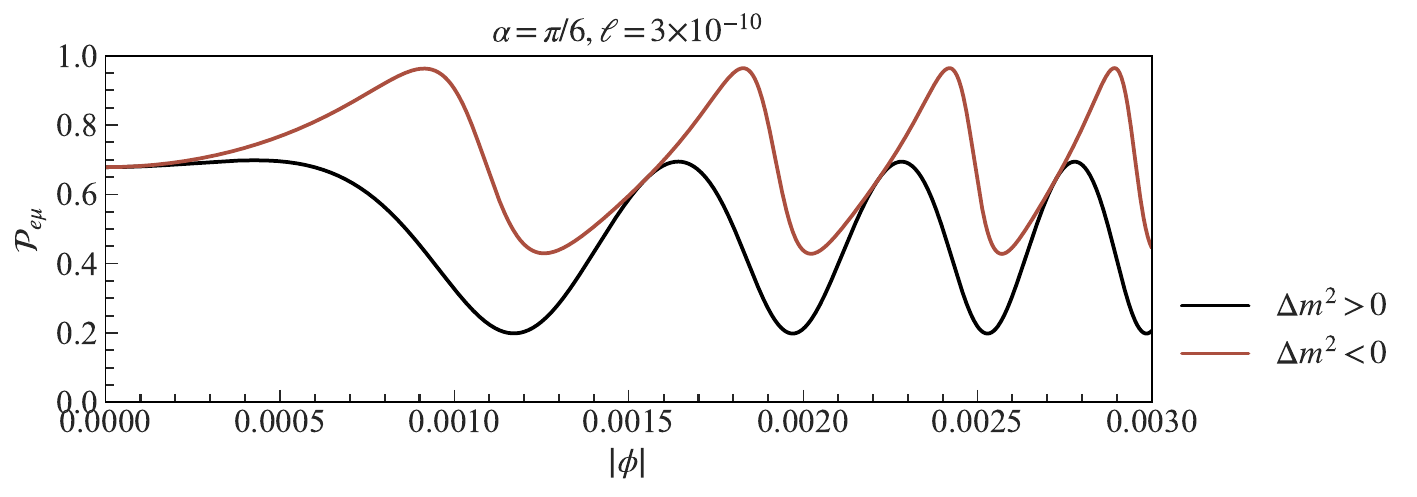}
\caption{\label{fig:prob1} The transition probability $\nu_e \to \nu_\mu$ is examined with respect to variations in the azimuthal angle $\varphi$, focusing on two specific values of the Lorentz--violating parameter: $\ell = 1\times10^{-10}$ and $\ell = 3\times10^{-10}$. The analysis is conducted within the framework of two--flavor neutrino oscillations, incorporating both normal and inverted mass hierarchies. Two different mixing angles, $\alpha = \pi/5$ and $\alpha = \pi/6$, are also considered to explore how the conversion behavior responds to changes in the oscillation parameters.}
\end{figure*}

\begin{figure*}
\centering
\includegraphics[height=5cm]{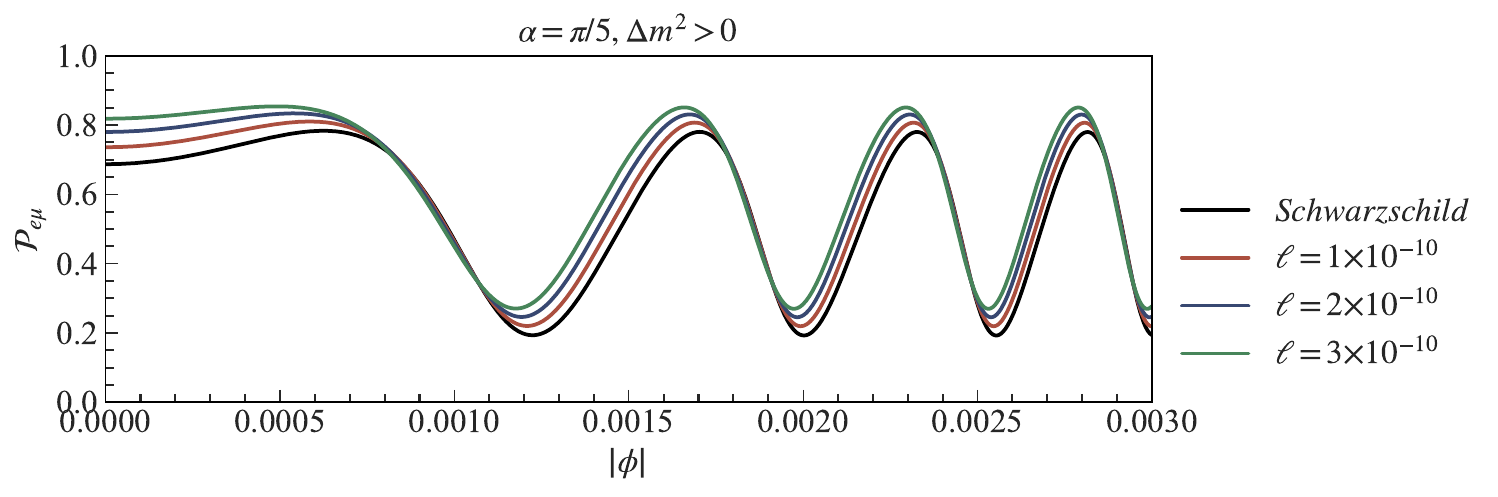}
\includegraphics[height=5cm]{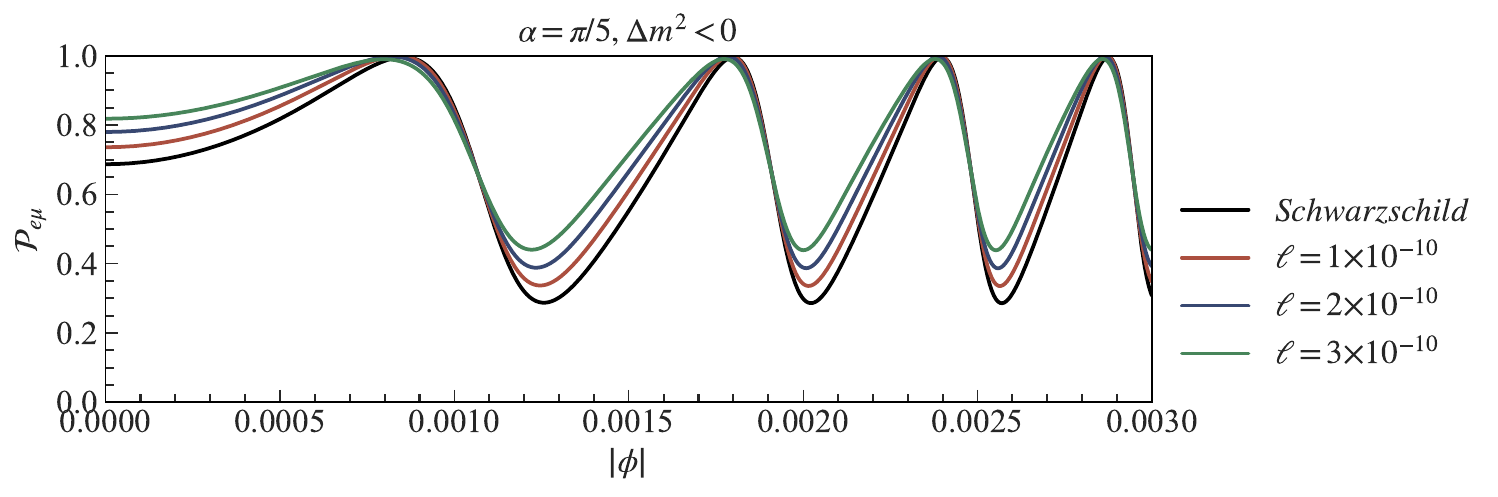}
\includegraphics[height=5cm]{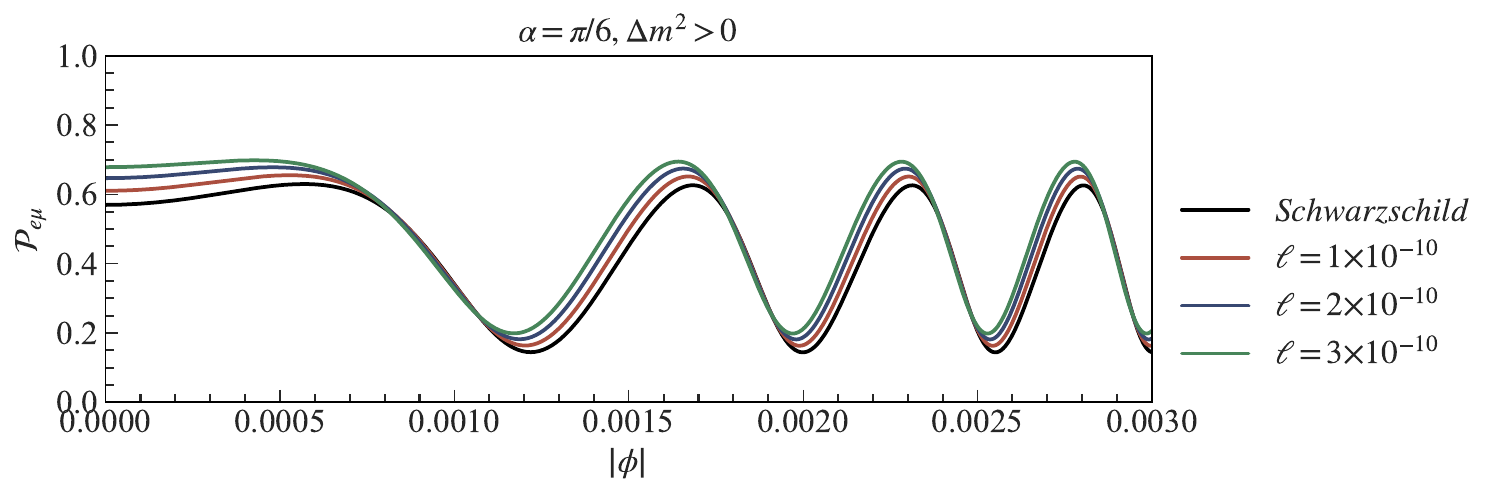}
\includegraphics[height=5cm]{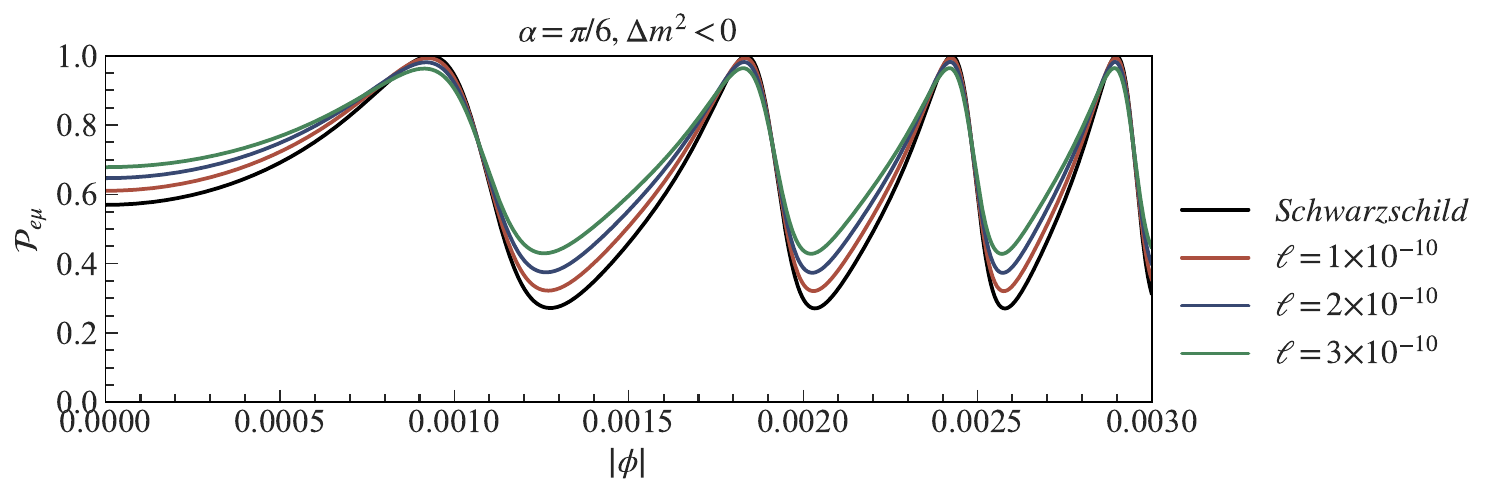}
\caption{\label{fig:prob2} The flavor conversion probability $\nu_e \to \nu_\mu$ is explored as it varies with the azimuthal angle $\varphi$, using different values of the Lorentz--violating parameter, $\ell = 0, 1\times10^{-10}, 2\times10^{-10}$ and $3\times10^{-10}$. This study is performed within the context of a two-flavor oscillation framework, incorporating both normal and inverted mass hierarchies. The influence of the mixing angle is examined by selecting $\alpha = \pi/5$ and $\alpha = \pi/6$ as benchmark values.}
\end{figure*}

\begin{figure*}
\centering
\includegraphics[height=5cm]{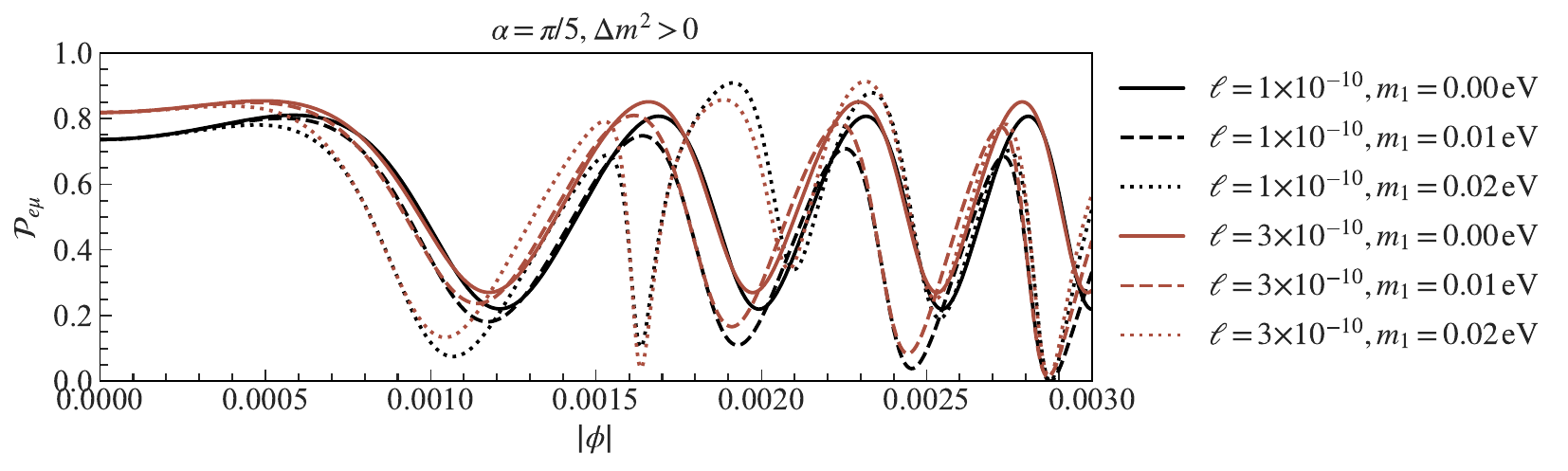}
\includegraphics[height=5cm]{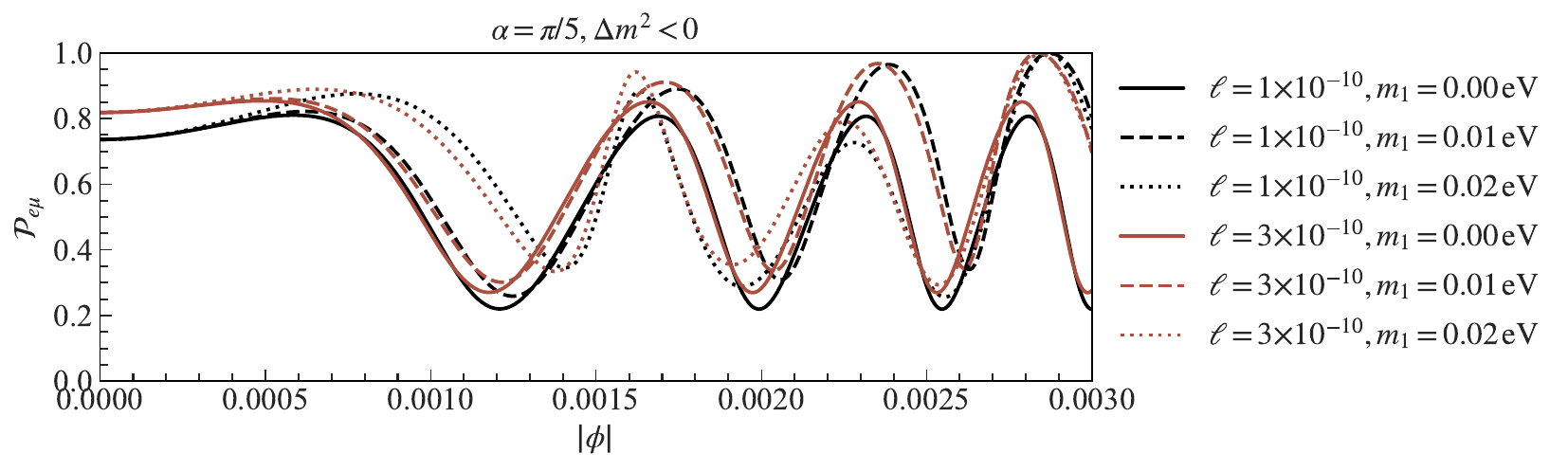}
\includegraphics[height=5cm]{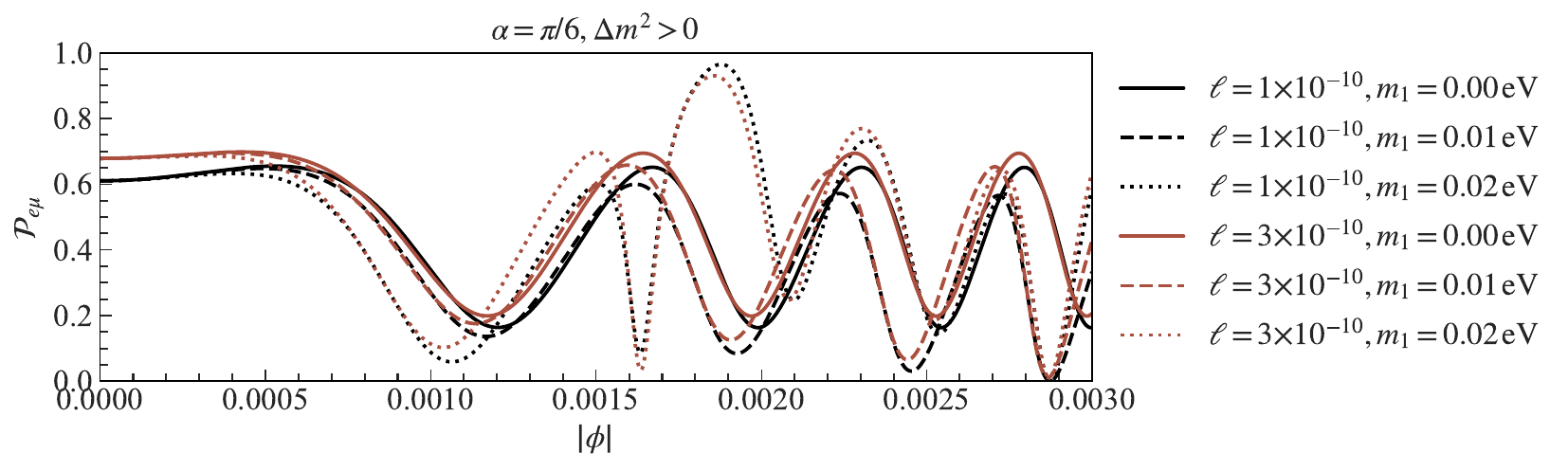}
\includegraphics[height=5cm]{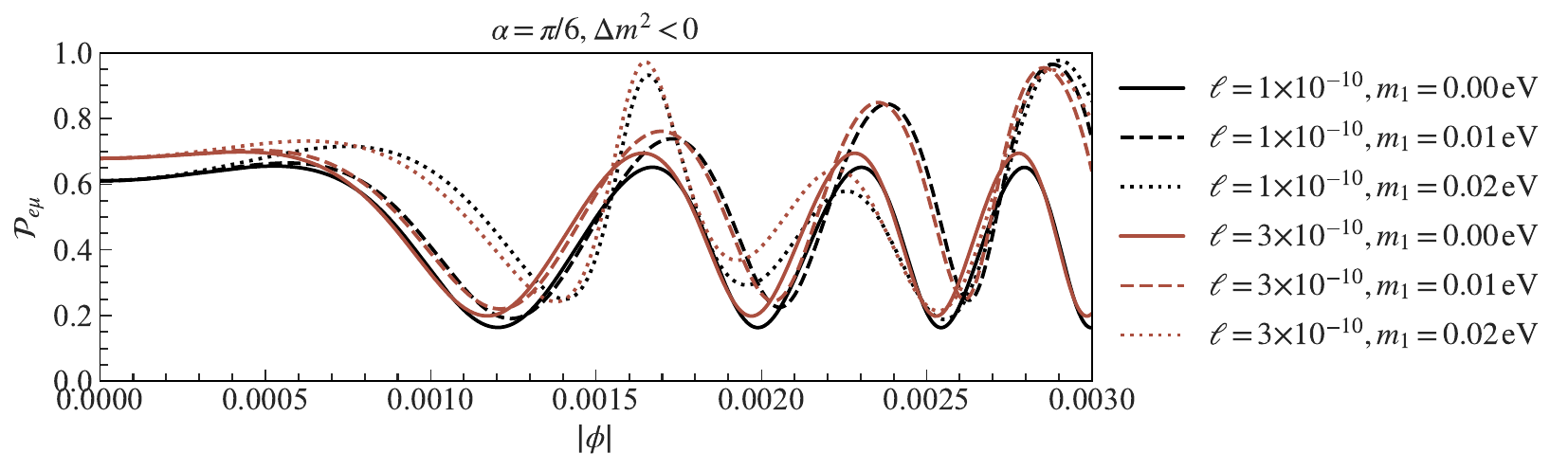}
\caption{\label{fig:prob3} The oscillation probability of neutrinos is plotted as a function of the azimuthal angle $\varphi$ for both normal ($\Delta m^2 > 0$) and inverted ($\Delta m^2 < 0$) mass hierarchies. The impact of Lorentz violation is illustrated through two parameter values: $\ell = 1\times10^{-10}$ (black curves) and $\ell = 3\times10^{-10}$ (red curves). Variations in the lightest neutrino mass are represented by different line styles: solid lines correspond to $m_1 = 0\,\mathrm{eV}$, dashed lines to $m_1 = 0.01\,\mathrm{eV}$, and dotted lines to $m_1 = 0.02\,\mathrm{eV}$.}
\end{figure*}

We analyzed how neutrino oscillations are affected by the spacetime of a Bumblebee black hole, focusing on gravitational corrections introduced through the Lorentz--violating parameter $\ell$, as we mentioned earlier. Assuming a circular trajectory of the detector around the Sun, where $x_D=r_D\cos\varphi$ and $y_D=r_D\sin\varphi$, we may numerically solve the quartic polynomial presented in Eq.~\eqref{solve_b} inside the equatorial plane, yielding two positive real roots, $b_1$ and $b_2$, for each $\varphi$. Each term in Eq.~\eqref{solve_b} is delineated by Eqs.~\eqref{Delta_m}-\eqref{sum_m}, with conclusions derived by replacing the specified values of $m_1$, $m_2$, $b_1$ and $b_2$. Oscillation patterns were compared between the Bumblebee scenario ($\ell \neq 0$) and the Schwarzschild case ($\ell = 0$), with Figs.~\ref{fig:prob1}, \ref{fig:prob2}, and \ref{fig:prob3} highlighting how the azimuthal angle $\varphi$ and the mass--squared difference $\Delta m^2$ influence flavor transition probabilities.

Moreover, Fig.~\ref{fig:prob1} shows that $\nu_e \rightarrow \nu_\mu$ oscillations are strongly affected by the mass ordering in Bumblebee spacetime. In the region $\varphi \in [0, 0.003]$, the transition probability is lower for the normal hierarchy ($\Delta m^2 > 0$) than for the inverted one ($\Delta m^2 < 0$). Adjusting the mixing angle from $\alpha = \pi/5$ to $\pi/6$ increases the peak probability for the inverted case while decreasing it for the normal one, indicating enhanced sensitivity of the mixing parameters to the background geometry.

The influence of $\ell$ is detailed in Fig.~\ref{fig:prob2}. Increasing $\ell$ from $1 \times 10^{-10}$ to $3 \times 10^{-10}$ slightly changes the oscillation frequency but noticeably increases the whole probability of conversion. This aspect results from a promotion in the effective phase due to radial stretching of the spacetime. It is worthy to be highlight that, when the Kalb--Ramond fields takes into account instead, it did not produce such radial rescaling, showing no comparable suppression \cite{Shi:2025rfq}.

On the other hand, Fig.~\ref{fig:prob3} addresses the role of the absolute mass scale. As the lightest neutrino mass $m_1$ increases from $0$ to $0.02\,\mathrm{eV}$, the amplitude decreases for the normal hierarchy and increases for the inverted one—behavior also seen in Kalb--Ramond gravity. However, the two models differ at $\varphi = 0$ due to distinct modifications of the metric: Bumblebee spacetime preserves $\mathfrak{g}_{tt}$, whereas Kalb--Ramond modifies it via a $1/(1-\ell)$ factor, changing local energy measurements and initial oscillation conditions.

The differences between Bumblebee and Kalb--Ramond black holes stem from how each theory breaks Lorentz symmetry and modifies spacetime. The Bumblebee model alters only $\mathfrak{g}_{rr}$ via spontaneous vector field symmetry breaking, while Kalb--Ramond affects both $\mathfrak{g}_{tt}$ and $\mathfrak{g}_{rr}$ through coupling to an antisymmetric tensor. This leads to different phase structures: Bumblebee affects the path length, and Kalb--Ramond alters both path and energy measurements. The Bumblebee field's anisotropic corrections enhance sensitivity to the sign of $\Delta m^2$ through interference with neutrino mass eigenstates. As a result, Bumblebee spacetime produces azimuth--dependent oscillation features and greater mixing angle sensitivity, while Kalb--Ramond leads to more uniform modulation. These contrasting behaviors reflect how vector--field symmetry breaking and antisymmetric--field renormalization shape flavor evolution—providing a potential background to distinguish quantum gravity effects in astrophysical neutrino scenarios.

Furthermore, as shown in the previous sections of this paper, the Lorentz--violating vector field of bumblebee gravity alters only the radial metric component $g_{rr}$, effectively stretching path lengths and increasing the oscillation phase; this leads to $\sim20\%$ changes in the $\nu_e\!\to\!\nu_\mu$ conversion amplitude relative to the Schwarzschild case and a hierarchy--dependent modulation that reaches $\gtrsim12\%$ across the azimuthal range.  Such differences become observationally relevant in environments where neutrinos graze a compact object: (i) a core--collapse supernova whose proto–neutron star promptly forms a black-hole remnant, (ii) the hyper--accreting disk left after a binary--neutron‐star merger--both bright MeV sources that Hyper--Kamiokande or DUNE would record with $\mathcal{O}(10^{5})$ events at 10 kpc, yielding sub-percent statistical errors, and (iii) TeV--PeV neutrinos emitted along blazar jets, whose trajectories skirt the central super-massive black hole and are detectable by IceCube--Upgrade.  Because the predicted flavor ratios differ from standard oscillations by $\sim10\!-\!20\%$, a net experimental accuracy of $\lesssim5\%$ in flavor identification suffices for a $3\sigma$ test; this benchmark lies within the projected capabilities of JUNO, DUNE and Hyper--K for MeV energies and IceCube–Gen2 for the high-energy channel.

\section{Conclusion}

This work examined how spontaneous Lorentz--symmetry breaking, encoded by parameter $\ell$, in Bumblebee gravity, reshaped every facet of neutrino behavior in the vicinity of a Schwarzschild--like black hole. By rebuilding the standard formalism for the annihilation channel $\nu\bar\nu\!\to\!e^{+}e^{-}$ on the modified metric, the study showed that the total energy deposition rate increased as $\ell$ grew: for a representative source of radius $R=20\,\mathrm{km}$, and luminosity $10^{53}\,\mathrm{erg/s}$ measured at infinity, the integrated output climbed by about five percent at $\ell = 0.1$ and by roughly fourteen percent at $\ell = 0.3$, while preserving the same radial profile as in general relativity. Such increase features arose because the vector field altered both the radial component $\mathfrak{g}_{rr}$ and subsequently the gradient of the differential rate $\mathrm{d}\dot Q/\mathrm{d}r$.

Propagation effects were governed by a reduced oscillation phase
$\Tilde{\Phi}_k=\frac{m_k^2}{2E_0}\left(1+\frac{\ell}{2}\right)(r_D+r_S)\left(1-\frac{b^2}{2r_Dr_S}+\frac{2M}{r_D+r_S}\right)$,
so higher values of $\ell$ lengthened the oscillation scale along both radial and lensed paths.  Numerical simulations revealed that raising $\ell$ from $1\times10^{-10}$ to $3\times10^{-10}$ left the azimuthal period almost unchanged but suppressed the $\nu_e\to\nu_\mu$ transition amplitude by up to twenty percent, an effect traced to the radial stretching of the optical metric. We pointed out that the increment of the oscillation amplitude was the relevant observable. For neutrinos coming from distant astrophysical sources, the rapid oscillations were averaged out, making the phase unobservable. Still, the increased amplitude in the flavor conversion probability, produced by Lorentz violation, remained as a stable and potentially measurable effect.

More so, the inverted mass ordering consistently produced higher conversion probabilities than the normal one with changing the lightest mass from $0$ to $0.02\,\mathrm{eV}$. For phase calculations, we employed Schwarzschild-like coordinates for coordinate dependency, which hold true at the astrophysical scales utilized ($r_S=10^5r_D\gg2M$). Near-event-horizon effects ($r\to2M$) were not included in these parameter selections. The phase definition itself was geometrically invariant and independent of particular coordinate choices, as Ref.~\cite{neu54} thoroughly demonstrated. The authors highlighted that the metric's adjustments to actual distance and energy were what caused changes in the oscillation phase, not the fictitious idea of a "gravitational phase". We think that extreme conditions, including neutron star merging events, can detect the oscillation alterations suggested in Figs.~\ref{fig:prob1}-\ref{fig:prob3}. Our numerical simulations have investigated the region of $\ell\geq10^{-10}$, where IceCube-Gen2's sensitivity may be able to detect Lorentz symmetry breaking signals.

Contrasting the present results with those obtained for Kalb--Ramond gravity highlighted the role of the symmetry--breaking sector.  Whereas the antisymmetric tensor field deepened the potential and curtailed energy deposition, the Bumblebee vector field acted effectively repulsively. Moreover, because the Bumblebee background preserved $\mathfrak{g}_{tt}$ (in comparison with the Schwarzschild case), lensing--induced oscillation patterns depended chiefly on path--length variations rather than on redshifted energies, yielding azimuth--dependent features and enhanced mixing--angle sensitivity that were absent in the tensor--driven model. 

In the future, more realistic techniques will need to incorporate quantum field theory, even though the toy model presented in this study can be considered a first step. Moreover, gravitational spin-flip transitions of neutrinos and gravitational decoherence effects must be taken into account in order to completely comprehend how gravitational influences affect neutrino oscillations. Although the consequences of quantum decoherence may be overlooked in the weak lensing limit, as noted in Ref.~\cite{neu53}, it is a constraint for neutrino long-range transmission. Detectors need to be able to regulate wave packets with great accuracy when the lensing effect produces mass-dependent oscillation modes.

Therefore, in a general panorama, the analysis showed that spontaneous Lorentz violation in Bumblebee gravity enhanced the energy released by neutrino processes, extended the oscillation lengths, and introduced distinct modulations in the flavor transition probabilities that depend on the mass ordering.


\section*{Acknowledgments}
\hspace{0.5cm} A. A. Araújo Filho is supported by Conselho Nacional de Desenvolvimento Cient\'{\i}fico e Tecnol\'{o}gico (CNPq) and Fundação de Apoio à Pesquisa do Estado da Paraíba (FAPESQ), project No. 150891/2023-7.

\bibliographystyle{ieeetr}
\bibliography{main}

\end{document}